\documentclass[a4paper,twoside]{article}

\usepackage{url}
\usepackage{epsfig}
\usepackage{subfigure}
\usepackage{calc}
\usepackage{amssymb}
\usepackage{amstext}
\usepackage{amsmath}
\usepackage{multicol}
\usepackage{pslatex}
\usepackage{apalike}
\usepackage{fancyhdr}
\usepackage{insticc}
\usepackage{graphicx}
\usepackage[small]{caption}

\begin{document}
\onecolumn \maketitle \normalsize \vfill

\section{\uppercase{Introduction}}\label{sec:introduction}

\noindent Source-filter modeling is one of the most widely used in speech processing.
Its success is certainly due to the physiological interpretation it relies on.
In this approach, speech is considered as the result of a glottal flow filtered by the vocal tract cavities and radiated by the lips.
Our paper focuses on the glottal source estimation directly from the speech signal.
Typical applications where this issue is of interest are voice quality assessment, statistical parametric
speech synthesis, voice pathologies detection, expressive speech production,... 

The goal of this paper is twofold. First a simple principle based on anticausality domination is presented.
Secondly, different source estimation techniques are compared according to their robustness. Their decomposition
quality is assessed in different conditions via two objective criteria : a spectral distortion measure and a
glottal formant determination rate. Robust source estimation is of a paramount importance since
final applications have to face adverse decomposition conditions on real continuous speech.

The paper is structured as follows. In section \ref{sec:theory} a theoretical
background on source estimation methods is given. The experimental protocol
we used for the comparison is defined in Section \ref{sec:protocol}.
In Section \ref{sec:results} results are exposed and the impact of different factors
on the estimation quality is discussed. Section \ref{sec:conclusion} concludes the paper
and proposes some guidelines for future work.

%%%%%%%%%%%%%%%%%%%%%
\section{\uppercase{Source estimation techniques}}\label{sec:theory}

\begin{figure}
	\centering
		\includegraphics[width=0.5\textwidth]{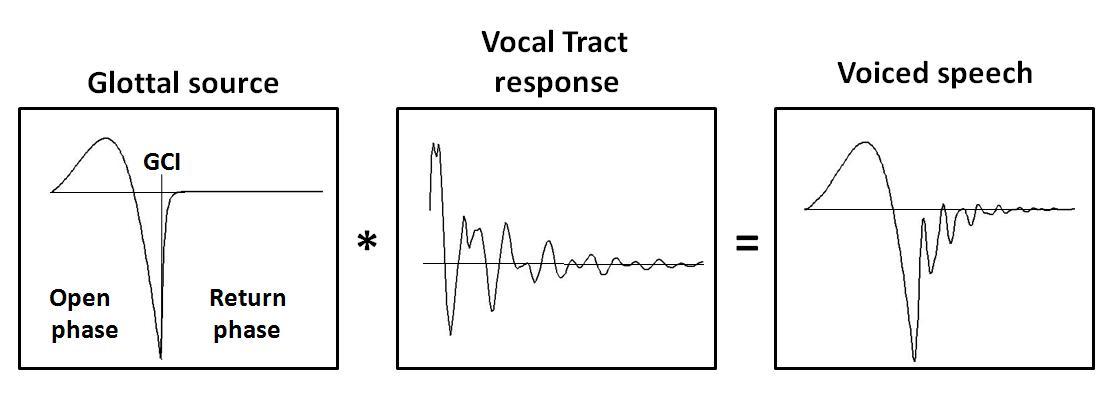}
		\caption{Illustration of the source-filter modeling for one voiced period. The Glottal Closure Instant (GCI) has the particularity
		to allow the separation of glottal open and closed phases, corresponding respectively to anticausal and causal signals. 
		}
	\label{fig:Source-Filter}
\end{figure}

\noindent We here present two popular voice source estimation methods, namely the
Zeros of the Z-Transform decomposition (ZZT) and the Iterative Adaptive Inverse Filtering technique (IAIF).
ZZT basis relies on the observation that speech is a mixed-phase signal \cite{Doval} where the anticausal component
corresponds to the vocal folds open phase, and where the causal component comprises both the glottis
closure and the vocal tract contributions (see Figure \ref{fig:Source-Filter}). As for the IAIF method, it isolates the source signal by iteratively
estimating vocal tract and source parts. After this brief state of the art, our approach based on Anticausality Dominated Regions (ACDR) is explained.

%%%%%%%%%%%%%%%%%%%%%
\subsection{ZZT-based decomposition of speech}

\noindent For a series of $N$ samples $(x(0),x(1),...x(N-1))$ taken from a
discrete signal $x(n)$, the $ZZT$ representation is defined as
the set of roots (zeros) $(Z_1,Z_2,...Z_{N-1})$ of the corresponding
Z-Transform $X(z)$:

\begin{equation}X(z)=\sum_{n=0}^{N-1} x(n)z^{-n}=x(0)z^{-N+1}\prod_{m=1}^{N-1} (z-Z_m)\label{eq:ZZT}\end{equation}

In order to decompose speech into its causal and anticausal
contributions \cite{Bozkurt}, ZZT are computed on frames centered
on each Glottal Closure Instant (GCI) and whose length is twice
the fundamental period at the considered GCI. These latter instants
can be obtained either by electroglottographic (EGG) recordings or by extraction methods
applied on the speech signal (see \cite{Kawahara} for instance).
The spectrum of the glottal source open phase is then computed from zeros outside the
unit circle (anticausal component) while zeros with modulus lower
than 1 give the vocal tract transmittance modulated by the source
return phase spectrum (causal component).

%%%%%%%%%%%%%%%%%%%%%
\subsection{Iterative Adaptive Inverse Filtering}
The inverse filtering technique aims at removing the vocal tract contribution from speech by filtering this signal by the inverse of an estimation of the vocal tract transmittance (this estimation being usually obtained by LPC analysis). Many methods implement the inverse filtering in an iterative way in order to obtain a reliable glottal source estimation.
\par
One of the most popular iterative method is the IAIF (Iterative Adaptive Inverse Filtering) algorithm proposed in \cite{Alku1992}. In its first version, this method implements LPC analysis so as to estimate the vocal tract response and use this estimation in the inverse filtering procedure. Authors proposed an improvement in \cite{Alku2000}, in which the LPC analysis is replaced by the Discrete All Pole (DAP) modeling technique \cite{ElJaroudi1991}, more accurate than LPC analysis for high-pitched voices.
\par
\begin{figure}[!htbp]
	\centering
	\includegraphics [width=.5\linewidth] {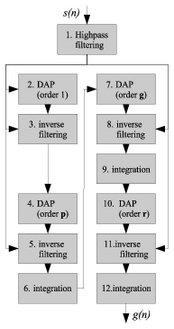}
	\caption{Block diagram of the IAIF method (from the documentation of TKK $Aparat$ \cite{Aparat2008}).}
	\label{fig:IAIF}
\end{figure}

The block diagram of the IAIF method is shown in Figure \ref{fig:IAIF} where $s(n)$ stands for the speech signal and $g(n)$ for the glottal source estimation.
The $1^{st}$ block performs a high-pass filtering in order to reduce the low frequency fluctuations inherent to the recording step. The $2^{nd}$ and $3^{rd}$ blocks compute a first estimation of the vocal tract, which is used in the $4^{th}$ and $5^{th}$ blocks to compute a first estimation of the glottal source. This estimation is the basis of the second part of the diagram ($7^{th}$ to $12^{th}$ blocks) where the same treatment is applied in order to obtain the final glottal source estimation.
\par
Based on this method the TKK $Aparat$ \cite{Airas2008} has been developed as a sofware package providing an estimation of the glottal source and its model-based parameters. We used the toolbox available on the TKK $Aparat$ website \cite{Aparat2008} for our experiments.

%%%%%%%%%%%%%%%%%%%%%
\subsection{Causality/anticausality Dominated Regions}

\begin{figure}
	\centering
		\includegraphics[width=0.5\textwidth]{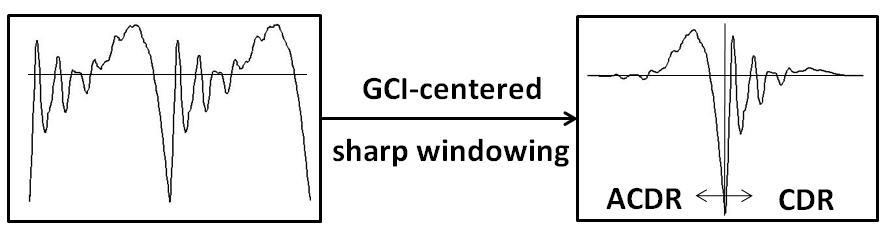}
		\caption{Effect of a sharp GCI-centered windowing on a two-period long speech frame. The Anticausality Dominated Region (ACDR) approximates
		the glottal source open phase.}
	\label{fig:Windowing}
\end{figure}

\noindent As previously mentioned, analysis is generally performed on
two-period long GCI-centered speech frames. Since GCI can be interpreted as
the starting point for both causal and anticausal phases, it dermarcates the boundary
between causality/anticausality dominated regions. As the domination zone of influence is limited
around the GCI, a sharp window (typically a Hanning-Poisson or Blackman window) is
applied to the analysis frame (see Figure \ref{fig:Windowing}). Since the causal
contribution (comprising the source return phase and the vocal tract components)
from the previous period is generally negligible just before the current GCI, the Anticausality Dominated Region (ACDR)
makes a good approximation of the source open phase. As long as the window is centered on a GCI and is sufficiently sharp,
this simple principle is applicable directly to the speech signal and even more on a first source estimation (obtained by IAIF for example). The dependency on the GCI detection for both ZZT and ACDR techniques will be discussed in Section \ref{ssec:GCI}. 

\section{\uppercase{Experimental Protocol}}\label{sec:protocol}

\begin{figure}
	\centering
		\includegraphics[width=0.5\textwidth]{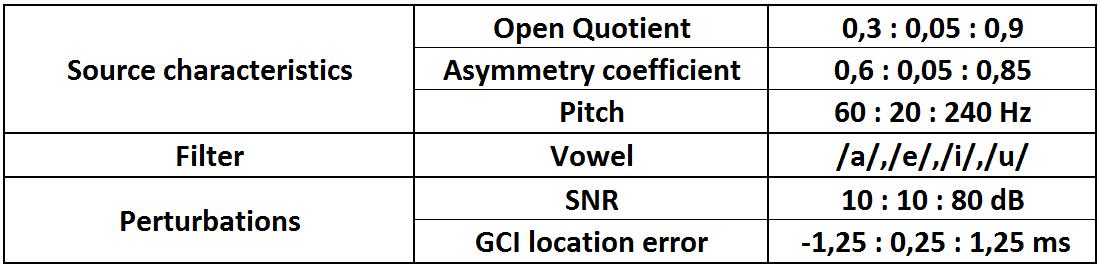}
		\caption{Table of parameter variation range.}
	\label{fig:Tableau}
\end{figure}

\noindent The experimental protocol we opted for is close to the one presented in \cite{Sturmel}.
Decomposition is achieved on synthetic speech signals for different test conditions.
The idea is to cover the diversity of configurations one could find in continuous speech
by varying all parameters over their whole range. Synthetic speech is produced according to the source-filter
model by passing a known train of Liljencrants-Fant glottal waves \cite{Fant} through an auto-regressive filter extracted
by LPC analysis on real sustained vowel uttered by a male speaker. As the mean pitch during these utterances was about 100 Hz,
it reasonable to consider that the fundamental frequency should not exceed 60 and 240 Hz in continuous speech.
Perturbations are modeled in two ways: by adding a white Gaussian noise on the speech signal and by making an error on the GCI location (see Sections \ref{ssec:noise} and \ref{ssec:GCI}). Figure \ref{fig:Tableau} summarizes all test conditions (which makes a total of 59280 experiments). 

Four source estimation techniques are here compared : ZZT, IAIF, ACDR principle applied to both speech and IAIF source frames.
In order to assess the decomposition quality we used two objective measures:

\begin{itemize}

\item {\bf Spectral distortion} : Many frequency-domain measures for quantifying the distance between two speech frames $x$ and $y$ arise from
the speech coding litterature. Ideally the subjective ear sensitivity should be formalised by incorporating psychoacoustic effects such
as masking or isophone curves. A simple relevant measure is the spectral distortion (SD) defined as:

\begin{equation}\label{eq:ZZT}
SD(x,y) = \sqrt{\int_{-\pi}^\pi(20\log_{10}|\frac{X(\omega)}{Y(\omega)}|)^2\frac{\text{d}\omega}{2\pi}}
\end{equation}

where $X(\omega)$ and $Y(\omega)$ denote both signals spectra in normalized angular frequency. In \cite{Paliwal},
authors argue that a difference of about 1dB (with a sampling rate of 8kHz) is rather imperceptible.
In order to have this point of reference between estimated and targeted sources we used the following measure:
\begin{equation}\label{eq:ZZT}
SD(x,y) \approx \sqrt{\frac{2}{8000}\sum_{20}^{4000}{(20\log_{10}|\frac{S_{estimated}(f)}{S_{reference}(f)}|)^2}}
\end{equation}

\item {\bf Glottal formant determination rate} : The amplitude spectrum for a voiced source (as shown in Figure \ref{fig:Source-Filter})
generally presents a resonance called \emph{glottal formant}. As this latter parameter is an essential feature, an error
on its determination after decomposition should be penalized. An example of relative error on the glottal formant determination is displayed
in Figure \ref{fig:histogram} for $SNR=50dB$. Many attributes characterizing a histogram can be proposed to evaluate a technique performance.
The one we used for our results is the underlying surface between $\pm10\%$ of relative error, which is an image of the determination rate given these bounds.

\begin{figure}
	\centering
		\includegraphics[width=0.5\textwidth]{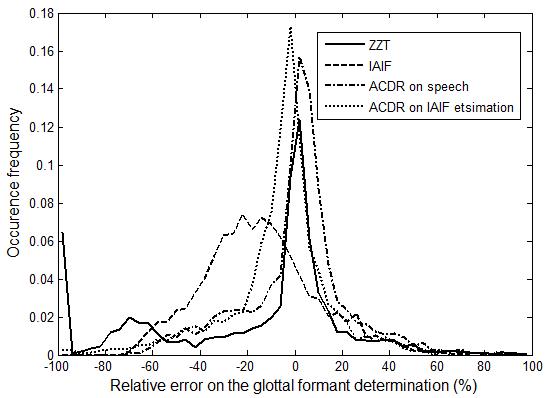}
		\caption{Histogram of relative error on the glottal formant determination (SNR=50dB).}
	\label{fig:histogram}
\end{figure}

\end{itemize}

\noindent In the next Section results are averaged for all considered frames.

%%%%%%%%%%%%%%%%%%%%%%%%%%%%%%%%%%%%%%%%%%%%%%%%%%%%%%%%%%%%
\section{\uppercase{Results}}\label{sec:results}

\noindent A quantitative comparison between described methods is here presented. More precisely results are
oriented so as to answer the two following questions: "\emph{How techniques are sensitive to perturbations such as noise or GCI location error?}" and "\emph{What is the impact of factors such as the fundamental frequency or the first formant on the decompostion quality?}".

%%%%%%%%%%%%%%%%%%
\subsection{Noise sensitivity}\label{ssec:noise}
\begin{figure}
	\centering
		\includegraphics[width=0.4\textwidth]{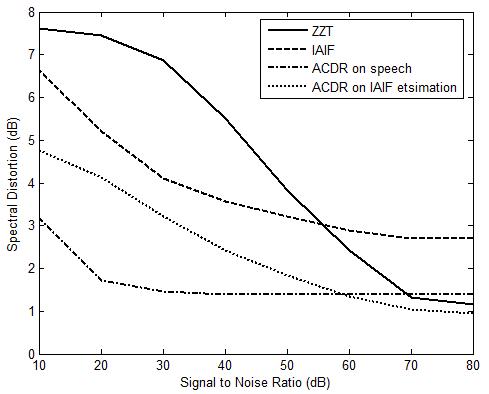}
		\caption{Impact of noise on the spectral distortion.}
	\label{fig:SNRonSpecError}
\end{figure}

\begin{figure}
	\centering
		\includegraphics[width=0.4\textwidth]{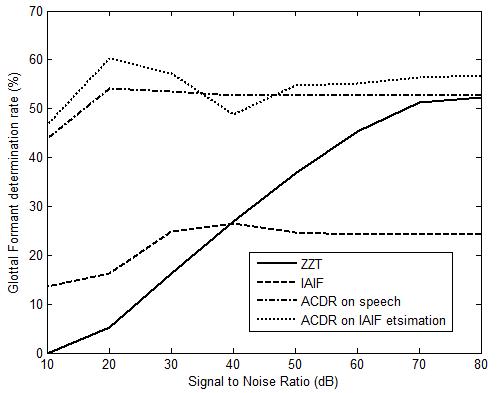}
		\caption{Impact of noise on the glottal formant determination rate.}
	\label{fig:SNRonFg}
\end{figure}

\noindent As a reminder a white Gaussian noise has been added to the speech signal at different SNR levels. This noise models not only recording or production noise but also every little deviation to the theoretical framework which distinguishes real and synthetic speech. Results according to both spectral distortion and glottal formant determination rate are displayed in Figures \ref{fig:SNRonSpecError} and \ref{fig:SNRonFg}. Among all techniques, ZZT turns out to be the most sensitive. This can be explained by the fact that a weak presence of noise may dramatically perturb the roots position in the Z-plane, and consequently the decomposition quality. Interestingly the utility of applying our proposed ACDR concept is clearly highlighted (see notably the improvement when applied to the IAIF estimation). Even when directly performed on the speech signal, ACDR principle clearly yields robust and efficient results.

%%%%%%%%%%%%%%%%%%
\subsection{GCI location sensitivity}\label{ssec:GCI}
\begin{figure}
	\centering
		\includegraphics[width=0.4\textwidth]{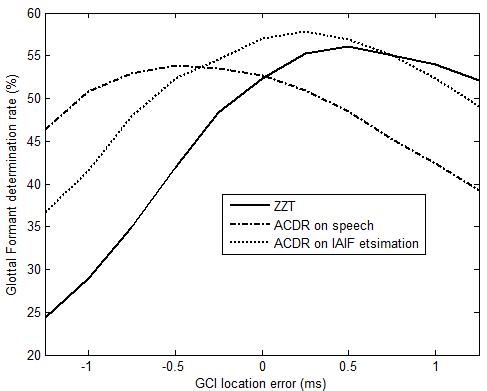}
		\caption{Impact of a GCI location error on the glottal formant determination rate (clean conditions).}
	\label{fig:GCIonFg}
\end{figure}

\noindent Another perturbation that could affect a method accuracy is a possible error made on the GCI location. Detecting these particular events directly on speech with a reliable precision is still an open problem although some interesting ideas have been proposed \cite{Kawahara}. Consequently it is rare that detected GCIs exactly match their ideal position when analyzing real speech. To take this effect into account we have tested the influence of a deviation to the real GCI location (GCIs are known for synthetic signals). Results for the glottal formant determination rate are shown in Figure \ref{fig:GCIonFg} for clean conditions (no noise added). As mentioned in \cite{Bozkurt}, the ZZT technique is strongly sensitive to GCI detection, since this latter pertubation may affect the whole zeros computation. A similar performance degradation is also observed for ACDR-based methods due to their inherent way of operating. Nevertheless this effect occurs to a lesser extent.

%%%%%%%%%%%%%%%%%%%
\subsection{The influence of pitch}

\begin{figure}
	\centering
		\includegraphics[width=0.4\textwidth]{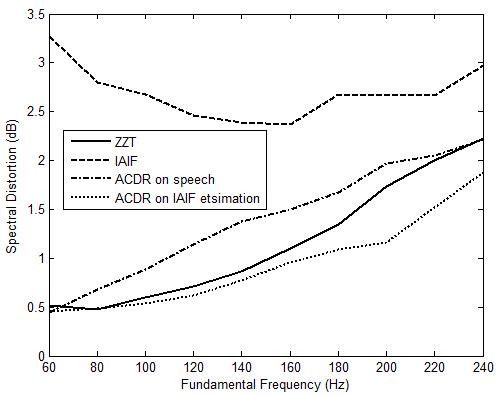}
		\caption{Impact of the fundamental frequency on the spectral distortion (clean conditions).}
	\label{fig:F0onSpecError}
\end{figure}

\noindent Female voices are known to be especially difficult to analyze and synthesize. The main reason is their high fundamental frequency which implies to treat shorter periods. As a matter of fact the vocal tract response has not the time to freely return to its initial state between two glottal sollication periods. Consequently the performance of ACDR method applied to high-pitched speech will intrinsically degrade, as it relies on the assumption that the vocal tract response is negligible in the ACDR. This hypothesis turns out to be acceptable in a certain extent and might be reconsidered for high pitch values. Figure \ref{fig:F0onSpecError} presents the evolution of spectral distortion with respect to the fundamental frequency. Unsurprinsingly all methods degrade as the pitch increases, and this in a comparable way.

%%%%%%%%%%%%%%%%%%%%
\subsection{The influence of the first formant}

\noindent In \cite{Bozkurt2004}, authors already reported erroneous glottal formant detection due to incomplete separation of $F_1$. As argued in previous subsection, particular configurations may lead to reconsider the assumption of ACDR applied on speech. More precisely, decomposition quality mainly depends on the 3 following parameters relative values: the pitch ($F_0$), the first formant ($F_1$) and the glottal formant ($F_g$). The greater is $F_0$ with regard to $F_1$ and $F_g$ and the more severe will be the decomposition conditions. Intuitively this latter case can be interpreted as an ever increasing interference between causal and anticausal parts.

\begin{figure}
	\centering
		\includegraphics[width=0.4\textwidth]{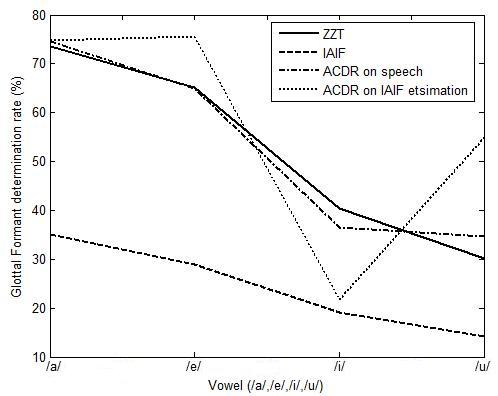}
		\caption{Impact of the first formant on the glottal formant determination rate (clean conditions).}
	\label{fig:VowelonFg}
\end{figure}

\noindent In our experiments filter coefficients were extracted by LPC analysis on four sustained vowels. Even though the whole spectrum may affect the decomposition, it is reasonable to consider that the effect of the first formant is preponderant. To give an idea, here are the corresponding first formant values: \emph{/a/:728Hz , /e/:520Hz , /i/:304Hz , /u/:218Hz}. The impact of the vowel on the decomposition accuracy is plotted in Figure \ref{fig:VowelonFg}. As expected a clear tendency of performance reduction as $F_1$ diminishes is observed.

%%%%%%%%%%%%%%%%%%%%%%%%%%%%%%%%%%%%%%%%%%%%%%%%%%%%%%%%%
\section{\uppercase{Conclusion and future work}}\label{sec:conclusion}

\noindent This paper addressed the problem of source estimation robustness. A comparison between four different techniques was carried out on a complete set of synthetic signals. These latter methods were the Zeros of the Z-Transform (ZZT), the Iterative Adaptive Inverse Filtering (IAIF), and our proposed concept of Anticausality Dominated Region (ACDR) applied either directly on speech, or on a first source estimation (thanks to IAIF in our case). Two formal criteria were used to assess their quality of decomposition: the spectral distortion and the glottal formant determination rate. Robustness was first evaluated by adding noise to speech. In a general way this noise modeled every little deviation to the ideal production scheme. Interestingly both ACDR-derived methods were the most robust and efficient. Another perturbation we considered was a possible error made on the GCI location. In a second step the influence of the pitch ($F_0$) and the first formant ($F_1$) was analyzed. Decomposition quality was interpreted as a trade-off between three amounts: $F_0$, $F_1$ and the glottal formant ($F_g$). In all our experiments ACDR-based techniques gave the more promising results. 

\noindent As future work we plan to investigate the incorporation of these methods in the following fields:

\begin{itemize}

\item {\bf Statistical parametric speech synthesis:} Hidden Markov models (HMM) have recently shown their ability to produce natural sounding speech \cite{Tokuda}. We already adapted this framework for the French language. A major drawback of such an approach is the "buzziness" of the generated voice. This inconvenience is typically due to the parametric representation of speech. Including a more subtle modeling of the voice source could lead to enhanced naturalness and intelligibility.

\item {\bf Expressive voice:} User-friendliness is one of the most important demand from the industry. Since expressivity is mainly managed by the source, an emotional voice synthesis engine should take into account realistic glottal source model parameters. Techniques presented in this paper could be used to estimate these parameters on speech samples extracted from an expressivity-oriented speech database.

\item {\bf Pathological speech analysis:} Speech pathologies are most of the time due to the irregular behaviour of the vocal folds during phonation. This irregular vibration can be induced by nodules or polyps on the folds and should result in irregular values of model parameters. Methods here presented could hence be used to estimate the glottal source and its features on pathological speech in order to quantify the pathology level.

\end{itemize}

\section*{\uppercase{Acknowledgments}}

\noindent Thomas Drugman is supported by the ``Fonds National de la Recherche
Scientifique'' (FNRS) and Nicolas D'Alessandro by the FRIA fundings.
The authors also would like to thank the Walloon Region for its
support (ECLIPSE WALEO II grant \#516009 and IRMA RESEAUX II grant
\#415911).

\renewcommand{\baselinestretch}{0.98}
\bibliographystyle{apalike}
{\small
\bibliography{example}}
\renewcommand{\baselinestretch}{1}

\end{document}